\documentclass[twocolumn,aps,prb,superscriptaddress]{revtex4}

\usepackage{amssymb}
\usepackage{amsmath}
\usepackage{graphicx}
\usepackage{amsfonts}

\begin{document}

\title{Compact cryogenic Kerr microscope for time-resolved studies\\ of electron spin transport in
microstructures}

\author{P. J. Rizo}
\affiliation{Zernike Institute for Advanced Materials, University of Groningen,\\
Nijenborgh 4, 9747 AG Groningen, The Netherlands}
\author{A. Pug\v{z}lys}
\affiliation{Zernike Institute for Advanced Materials, University of Groningen,\\
Nijenborgh 4, 9747 AG Groningen, The
Netherlands}
\affiliation{Photonics Institute, Vienna University of
Technology, Gusshausstrasse 27/387, 1040 Vienna, Austria}
\author{J. Liu}
\affiliation{Zernike Institute for Advanced Materials, University of Groningen,\\
Nijenborgh 4, 9747 AG Groningen, The Netherlands}
\author{D. Reuter}
\affiliation{Angewandte Festk\"{o}rperphysik, Ruhr-Universit\"{a}t
Bochum, D-44780 Bochum, Germany}
\author{A. D. Wieck}
\affiliation{Angewandte Festk\"{o}rperphysik, Ruhr-Universit\"{a}t
Bochum, D-44780 Bochum, Germany}
\author{C. H. van der Wal}
\affiliation{Zernike Institute for Advanced Materials, University of Groningen,\\
Nijenborgh 4, 9747 AG Groningen, The Netherlands}
\author{P. H. M. van Loosdrecht}
\affiliation{Zernike Institute for Advanced Materials, University of Groningen,\\
Nijenborgh 4, 9747 AG Groningen, The Netherlands}

\date{\today}

\begin{abstract}
A compact cryogenic Kerr microscope for operation in the small
volume of high-field magnets is described. It is suited for
measurements both in Voigt and Faraday configuration. Coupled with a
pulsed laser source, the microscope is used to measure the
time-resolved Kerr rotation response of semiconductor
microstructures with $\sim$1 micron spatial resolution. The
microscope was designed to study spin transport, a critical issue in
the field of spintronics. It is thus possible to generate spin
polarization at a given location on a microstructure and probe it at
a different location. The operation of the microscope is
demonstrated by time-resolved measurements of micrometer distance
diffusion of spin polarized electrons in a GaAs/AlGaAs
heterojunction quantum well at 4.2 K and 7 Tesla.
\end{abstract}

%\pacs{85.75.-d,72.25.Mk,75.70.Cn}
\maketitle

\section{Introduction}

The field of spintronics aims at manipulating the spin and charge
degrees of freedom of carriers in order to expand the
functionalities of current electronic devices. In this field,
microscopy based on the magneto-optical Kerr effect (Kerr
microscopy) is particularly valuable for measuring spin polarization
in ferromagnetic and semiconductor micro and nanostructures. In
recent years, Kerr microscopy has been applied by several research
groups to study phenomena relevant to spintronics. For example,
Awschalom and coworkers utilized a room-temperature optical
microscope, electromagnet and cryogenic cell with cold finger to
measure continuous-wave (cw) and time-resolved Kerr rotation in
biased semiconductor strips \cite{Kikkawa99}. Erskine and coworkers
studied room-temperature domain wall propagation in permalloy
nanowires subject to fast-rise time magnetic field step waveforms
utilizing an oblique-incidence Kerr polarimeter with 2 $\mu$m
spatial resolution and 1 ns temporal resolution
\cite{Beach05,Nistor06}. Imamo\u{g}lu and coworkers utilized a bath
cryostat with superconducting magnet that incorporated optics,
detectors and positioners inside the cryostat for cw measurements of
single spins in quantum dots \cite{Atature07}.

Characteristics that are desirable in a Kerr microscope for
spintronics research are the ability to apply large magnetic fields,
a time-resolution on the timescale of the spin dynamics, the ability
to excite and probe spins at different locations using different
photon energies and the possibility to operate in Voigt and Faraday
geometries. Application of large magnetic fields is often desirable
in order to produce large Zeeman splittings or to distinguish
between different spin populations in a sample. Time-resolving spin
dynamics has proven to be a very powerful tool in the field of
spintronics. It allows direct observation of spin precession,
dephasing and transport. Studies of spin transport, in particular,
benefit from the ability of exciting spin polarized carriers in one
place and detecting them at a different location. The possibility to
operate the microscope while applying the magnetic field parallel to
the sample plane (Voigt configuration) or perpendicular to the
sample plane (Faraday configuration) allows, for example,
controlling the magnetization orientation of ferromagnetic films
with perpendicular or in-plane magnetic anisotropy. Moreover, in
band-gap engineered semiconductor heterostructures, two-dimensional
charge carrier systems can be Zeeman- or Landau split in these
configurations, respectively. This can be exploited, for example, in
dilute magnetic semiconductors like GaN:Gd \cite{Lo08}.

Additional characteristics that are desirable in the Kerr microscope
are high sensitivity, high spatial resolution and possibility of
cryogenic operation. The sensitivity of the Kerr microscope is given
by the minimum Kerr rotation angle that can be measured and is
typically of the order of a few $\mu$rad. High spatial resolution
can be achieved with objective lenses of very short focal distance.
The spatial resolution \textit{r} is determined by the wavelength
$\lambda$ of the laser source utilized and the numerical aperture of
the objective lens \textit{NA}:

\begin{equation}\label{resolution}
r=\frac{1}{2}\cdot \frac{1.22\lambda}{NA}
\end{equation}

\noindent Submicrometer resolution is easily achievable using
high-\textit{NA} objective lenses (\textit{NA}$\sim$ 1) and laser
sources with wavelengths in the UV, visible or near IR range.
However, in practice, the small cryogenic measurement volumes inside
high-field magnets enforce long distances between room temperature
optics and the sample. This is in conflict with the convenience of
utilizing high-quality microscope objectives at room temperature.
The possibility of cryogenic operation is important, among other
reasons, because spin relaxation rates tend to be strongly
temperature dependent. This temperature dependence often helps
identifying the dominant spin relaxation mechanism.

In this article a cryogenic sample-scanning Kerr microscope for
static and time-resolved Kerr rotation (TRKR) measurements in high
magnetic fields is described. A compact design based on \textit{in
situ} operation of an aspherical singlet lens and a piezo-driven
translation stage allows obtaining micron-scale resolution inside
the bore of a large superconducting magnet. The instrument can
operate in Faraday and Voigt configurations and with femtosecond or
cw lasers. The spatial resolution can be as good as 1 micrometer.
Sample temperature can be set between 1.5 K and 300 K and magnetic
fields of up to 8 Tesla can be applied. The capabilities of the
microscope are demonstrated by measuring the time-resolved Kerr
rotation response of devices fabricated on a GaAs/AlGaAs
heterojunction quantum well at 4.2 K and 7 Tesla. The applicability
of the instrument for time-resolved studies of spin transport is
demonstrated by measuring micrometer distance diffusion of spins
precessing in an applied magnetic field.

\section{Instrument configuration}

\begin{figure}
\begin{center}
\includegraphics[width=7cm]{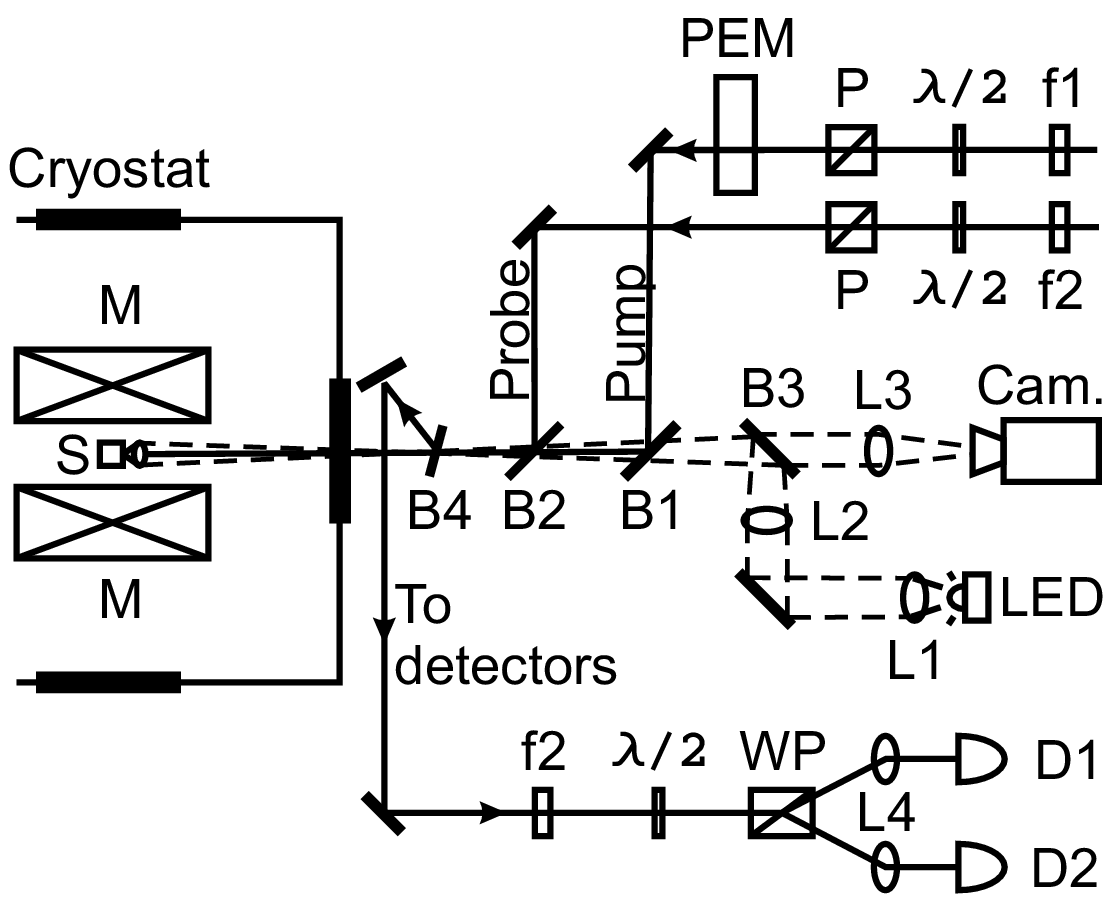}
\end{center}
\caption{Schematic of key components of the system. Pump and probe
beams are indicated. Each beam passes through different interference
filters f1 and f2. Half-wave plates ($\lambda/2$) and Glan-Thomson
polarizers (P) ensure proper beam polarizations. A
photoelastic-modulator (PEM) modulates the pump between right and
left circular polarizations at 50 kHz. Pump and probe beams are
steered using beamsplitters B1 and B2. The light from a white LED is
collected by lens L1 (focal length 4 cm) and focused by lens L2
(focal length 15 cm) a few cm in front of the objective lens in the
cryostat. Beamsplitter B3 sends the illumination light into the
objective lens. The sample is positioned inside the cryostat using
an XYZ piezo-driven translation stage (S). The cryostat is equipped
with a spit-coil superconducting magnet (M). Pump and probe beams as
well as the illumination light are focused onto the sample by the
objective lens in the cryostat which is depicted next to the
translation stage. Reflected laser and illumination light are
collimated by the objective lens and exit the cryostat. Beamsplitter
B4 collects the reflected laser light and sends it to the detection
system. Light from the sample surface passes through all
beamsplitters and is focused by lens L3 (focal length 19 cm) to form
an image on the camera (Cam.). In the detection system, interference
filter f2, rejects the pump light. A Wollaston prism (WP) separates
the reflected probe into s- and p-polarizations and a half-wave
plate is used to ensure equal intensities of both polarizations in
the absence of a pump pulse. Lenses L4 (focal length 10 cm) focus
the orthogonal polarizations of the probe light onto identical
photodiode detectors D1 and D2. } \label{schematic}
\end{figure}

The Kerr microscope consists of three main parts: 1. light source,
tabletop optics and detection system, 2. magneto-optical cryostat
and 3. insert which contains an XYZ sample positioner, objective
lens and electrical leads to the sample. Figure \ref{schematic}
shows a schematic of the most important components of the system. In
this section the configuration of the three parts of the microscope
will be discussed in detail, focusing on the configuration required
for time-resolved Kerr rotation measurements of spin transport in
semiconductor microstructures in the Voigt geometry. Data taken in
this configuration will be presented. The possibility of changing to
other configurations, for example to Faraday geometry or static Kerr
measurements, do not present major difficulties.

The light source, tabletop optics and detection system generate and
deliver the appropriate light beams to the sample and detect the
changes of polarization of the reflected beam. For time-resolved
Kerr rotation measurements, a pulsed laser system is utilized. In
this case two light pulses are required: a pump pulse which
generates a non-equilibrium spin polarization in the sample and a
probe pulse which measures this non-equilibrium spin polarization at
a certain location and at a different moment in time. Light pulses
are generated by a cavity-dumped Ti:sapphire laser (KMLabs Inc.
Cascade) with pulse width of 20 fs. The laser spectrum is centered
at 800 nm and has a full-width at half-maximum (FWHM) of $\sim$80
nm. A prism compressor is used to pre-compensate for dispersion
caused by several millimeters of calcite and quartz in the beam path
before the sample. A 1:5 beamsplitter is used to divide the single
beam from the laser source into separate pump and probe beams. The
time delay between the pump and probe pulses can be set by a
computer-controlled delay stage (Physik Instrumente M-531.DD) with
0.1 micron step size and travel range of 60 cm. Half-wave plates
($\lambda/2$ in Fig. \ref{schematic}) followed by Glan-Thomson
polarizers (P) in both pump and probe arms are used to produce beams
with well defined polarization. A photoelastic modulator (Hinds
PEM-90 I/FS50) placed on the path of the pump beam modulates its
polarization between right-circular polarization and left-circular
polarization at a frequency of 50 kHz. In order to study the laser
wavelength dependence of the Kerr response of the sample, 10 nm
($\sim$18 meV) FWHM spectral bandwidth interference filters are
placed in the path of both pump (f1) and probe beams (f2). The
resulting spectral filtering increases the width of both pulses to
approximately 120 fs. The measurements that will be discussed in the
following section were done utilizing filters centered at 780 nm for
the pump and 820 nm for the probe beams.

\begin{figure*}
\begin{center}
\includegraphics[width=15cm]{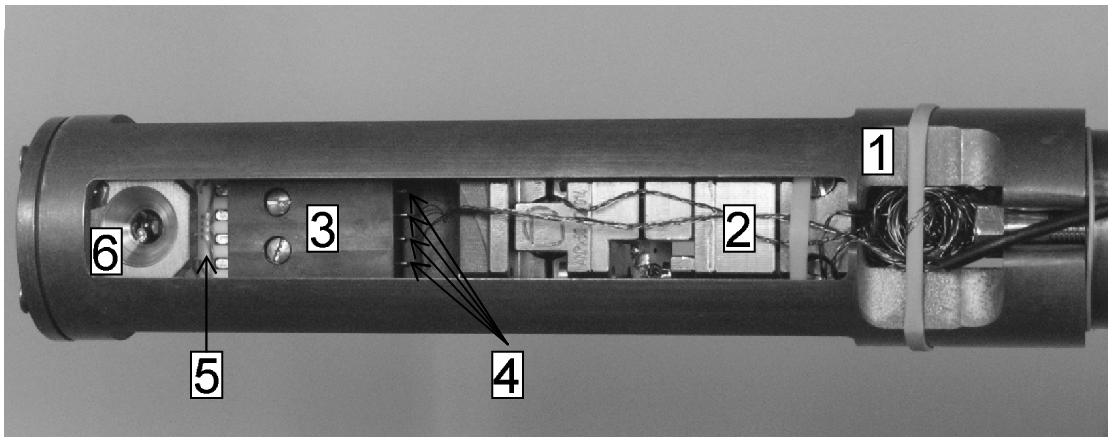}
\end{center}
\caption{Side view of the microscope insert. The parts of the insert
as discussed in the main text are: 1. copper housing, 2. XYZ
translation stage, 3. polyimide frame, 4. spring-loaded pins, 5.
modified ceramic chip carrier, 6. lens and lens holder. The
temperature sensor is on the back side and thus not visible.  }
\label{insert}
\end{figure*}

In order to control the positions of pump and probe beams on the
sample and their separation, imaging and beam steering optics were
installed. A near-IR sensitive camera (JAI CV-M50IR) is used for
imaging (Cam. in Fig. \ref{schematic}). Both imaging and
illumination are done through the objective lens in the microscope
insert (detail description of the microscope insert is given below).
The sample is illuminated using a collector lens (L1, focal length 4
cm) which collimates the light from a white light-emitting diode and
a condenser lens (L2, focal length 15 cm) which focuses this light a
few centimeters in front of the objective lens in the insert. The
objective lens focuses the white light illuminating an area on the
sample of approximately 150 microns in diameter. Light reflected
from the sample is collected by the objective lens and is focused
onto the camera using a large diameter achromatic lens with focal
length 19 cm (L3). The pump and probe beams incident on the sample
are also imaged by the camera allowing easy beam steering. The
position of pump and probe beams on the sample is manually
controlled by steering beamsplitters B1 and B2 using high-resolution
actuators. The angular sensitivity achievable in this configuration
is approx. 4.6 arcsecond. The effective focal length of the
objective lens is 2.75 mm, which gives a lateral beam displacement
sensitivity of $\sim$0.06 microns.

The detection system consists of a balanced photodiode bridge which
measures the rotation of the plane of polarization of the reflected
probe beam. The probe light reflected from the sample is collected
and collimated by the objective lens in the insert after which the
light is sent to the detection system by beamsplitter B4 indicated
in Fig. \ref{schematic}. An interference filter, f2 in the figure,
similar to the one in the probe beam path, is used to reject the
pump light. A half-wave plate rotates the probe polarization by
45$^\circ$ after which the probe is split into two beams with
\textit{s}- and \textit{p}-polarizations using a Wollaston prism
(WP). The emerging beams are detected by separate but identical
photodiodes D1 and D2. Voltage outputs from the photodiodes are
subtracted and sent to a lock-in amplifier. The voltage measured by
the lock-in amplifier at the frequency of the polarization
modulation (50 kHz) is, to first order, proportional to the Kerr
rotation angle. Kerr ellipticity contributes to the measured voltage
only as a third order correction. In order to balance the photodiode
bridge, the half-waveplate before the Wollaston prism is rotated
slightly until the difference between the signals measured with the
two photodiodes is minimum. A rotation of the plane of polarization
of the probe beam caused by the pump pulse causes an increase in the
intensity detected by one of the photodiodes and a decrease in the
intensity detected by the other. The difference between the two
intensities can be positive or negative, indicating Kerr rotation
angles of opposite signs.

A magneto-optical cryostat is used to control the sample temperature
and to apply a static magnetic field. The cryostat used was an
Oxford Instruments Spectromag SM4000-8 equipped with a split pair
superconducting split-coil magnet (M in Fig. \ref{schematic})
capable of producing magnetic fields of up to 8 Tesla. The sample is
cooled by He gas flow from a liquid He reservoir. Sample temperature
can be set between 1.5 K and 300 K. The split pair magnet allows
optical access to the sample with the magnetic field applied along
the optical axis (Faraday configuration) or perpendicular to it
(Voigt configuration). The orientation of the magnetic field can be
changed by 180$^\circ$ without modifying the optics by use of the
bipolar power supply of the magnet.

The sample insert is the most critical part of the Kerr microscope.
It contains the sample, objective lens, temperature sensor, sample
positioning and focusing system and electrical leads to the sample.
The insert is placed inside the magneto-optical cryostat, requiring
all parts to perform well at low temperatures and high magnetic
fields. A picture of the insert with all parts indicated is shown in
Fig. \ref{insert}. The insert consists of a cylindrical copper
housing fixed at the end of the sample rod supplied with the
magneto-optical cryostat. The cylinder diameter is 25 mm. The
objective lens is mounted on a copper/polyimide holder at the bottom
of the copper cylinder. The special lens holder ensures that no
stress is applied to the lens due to thermal contraction of the
materials during cool down. The objective lens will be discussed in
more detail later. The sample is mounted on a modified ceramic chip
carrier with eight electrical lead contacts and placed approximately
in the focal plane of the objective lens. The chip carrier is
mounted on an all-polyimide frame. The frame material was chosen to
have low thermal contraction and good electrical isolation
properties. Eight spring-loaded pins mounted on the frame make
electrical contact with the leads in the chip carrier. In addition,
the spring force of the pins holds the chip carrier in a fixed
position in the polyimide frame and allows changing the samples
easily by simply placing a different chip carrier into the frame. A
Cernox temperature sensor (LakeShore CX-1050-SD, not visible in the
figure) is placed on the frame in order to accurately monitor the
temperature of the sample.

The polyimide frame holding the sample is screwed onto an XYZ
inertial piezo-driven translation stage (Attocube Systems
ANPxyz50/LT) having 10 nm minimum step size at 4.2 K and a total
travel range of 4 mm in the x and y directions and 3 mm in the
vertical direction. The small step size of the translation stage
allows accurate scanning and positioning of microstructures with
respect to the focused laser beams while its long travel range
allows measuring several devices on a single sample\cite{Meyer05}.
After each step of the translation stage, the piezos go back to zero
voltage resulting in no drift of the sample position due to noise in
the piezo driving electronics. The translation stage holding the
polyimide frame and chip carrier is fixed inside of the copper
cylinder on the top part. This allows free displacement of the
sample with respect to the objective lens. Translating the sample
along the optical axis of the objective lens changes the focusing of
the laser beam as well as the focusing of the image of the device
surface captured by the camera. Translation in the remaining two
directions allows scanning the sample surface.

Proper selection of the objective lens in the insert is decisive to
the performance of the microscope. Since the microscope is designed
to work with a split-pair magnet both in Faraday and Voigt
configurations, space constrains inhibit the use of mulit-element
microscope objectives. Instead, a single, high-numerical aperture
aspherical lens has to be used.

Diffraction-limited focusing of the incoming laser beams ensures the
highest spatial resolution. Spherical and chromatic aberrations are
major obstacles in obtaining a diffraction-limited spot size on the
optical axis of the objective lens. For measurements of spin
transport it is necessary to ensure nearly diffraction-limited spots
not only on-axis but also off-axis. In this case it is essential to
minimize comatic aberration.

The detrimental effects of spherical aberration can be minimized by
utilizing specially designed aspheric lenses. A wide range of small
diameter high-NA aspheric lenses designed for wavelengths in the
vicinity of 800 nm are commercially available. In order to choose
the proper lens for the Kerr microscope geometrical ray tracing
calculations of selected aspheric lenses were performed. Incoming
parallel rays, both marginal and paraxial, were traced through the
lenses at different wavelengths within the laser spectrum.
Subsequently the diameter of the focal spot for each lens was
calculated when focusing the beams on-axis and off-axis. The
parameters that entered the calculation are the known refractive
index values of the lens material, the incoming light wavelengths
and the aspheric coefficients for both lens surfaces. A detailed
comparison was made for two aspheric lenses designed for 830 nm with
numerical aperture 0.68, one biconvex (LightPath 350330) and one
plano-convex (LightPath 350390). The calculations revealed that with
the biconvex lens on-axis diffraction-limited spot size can be
achieved at the design wavelength (spherical aberration-free). In
contrast, the plano-convex lens causes some spherical aberration
even at the design wavelength. Plano-convex lenses, however, have
minimum comatic and spherical aberrations when focusing a collimated
beam. In fact, comatic aberration of the biconvex lens was
calculated to be much larger than that of the plano-convex lens. By
taking this into account the plano-convex lens was chosen to be most
appropriate for the Kerr microscope. According to our calculations
for the plano-convex lens, the combined spherical and chromatic
aberrations (for 10 nm spectral bandwidth) increase the diameter of
the on-axis spot by only $\sim$6$\%$ compared to the
diffraction-limited case.

\begin{figure}
\begin{center}
\includegraphics[width=7cm]{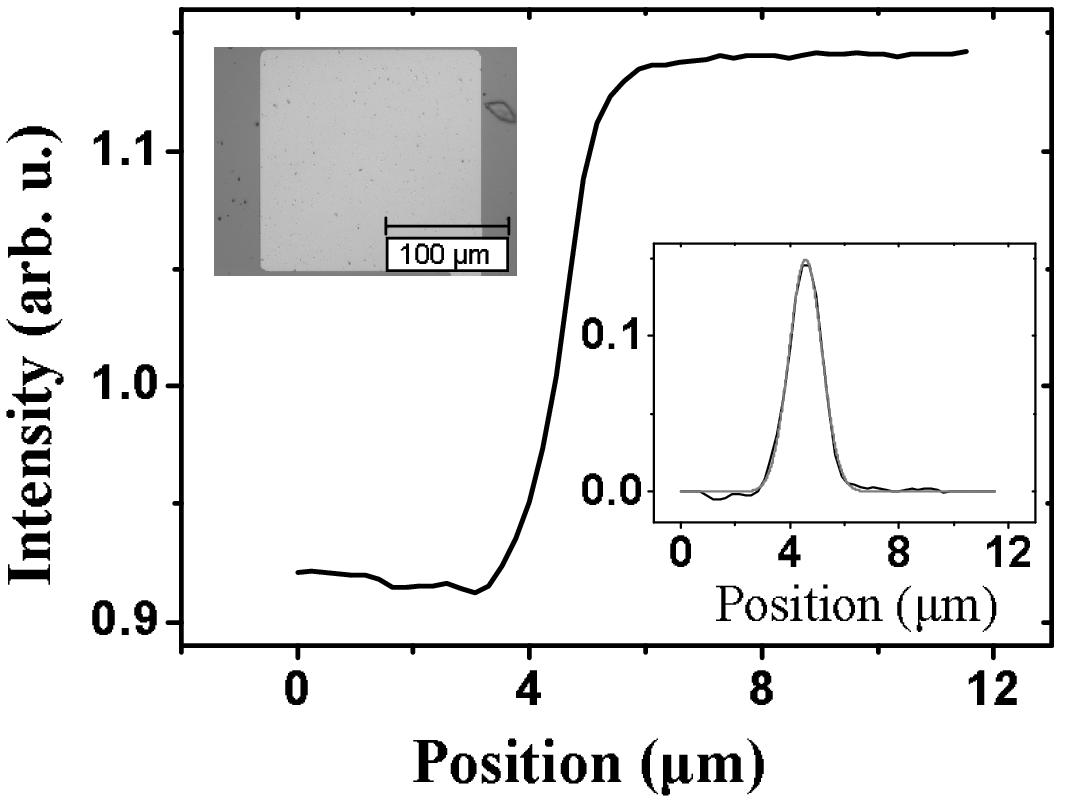}
\end{center}
\caption{Reflected intensity of the focused laser beam as a function
of lateral position of the edge of the gold pad. The derivative of
this trace, shown in the inset in the lower right corner, gives the
spatial intensity profile of the focused beam. The spot's FWHM at
4.2 K is $\sim$1.6 $\mu$m. The inset on the top-left corner shows an
optical image of the gold pad on the heterostructure surface. The
laser beam used in this measurements had a spectrum centered at 820
nm and spectral FWHM of 10 nm.} \label{spotsize}
\end{figure}

Spatial profiles of the laser light at the focus are measured with a
knife-edge technique. The laser beam is focused on the surface of
the sample, in this case, an epitaxially grown GaAs/AlGaAs
heterojunction structure. A square Ti/Au pad (170 $\mu$m wide and
150 nm thick) was deposited on the sample surface by lithography and
lift-off (see the inset on the top-left corner of Fig
\ref{spotsize}). The edge of the gold pad was used as knife-edge.
The laser beam is focused on the surface of the sample close to the
gold pad. The gold pad is oriented such that the sample can be
translated parallel and perpendicular to the edges of the square
pad. The reflection of the focused beam is collected and detected by
a silicon photodiode. The edge of the gold pad is moved across the
focused beam and the reflected light intensity is measured as a
function of the lateral displacement of the gold pad.

Utilizing the knife-edge technique, the focused spot profile was
measured at temperatures 4.2 K and 300 K for a laser beam with
spectrum centered at 820 nm and spectral FWHM of 10 nm. The
measurement at 4.2 K is presented in Fig. \ref{spotsize}. The
derivative of this trace, shown in the inset on the lower right
corner of the figure, is the spatial profile of the focused beam.
The focused spot profile has a nearly Gaussian shape with FWHM of
$\sim$1.6 $\mu$m. At 300 K the measured FWHM is $\sim$1 $\mu$m,
consistent with the spot diameter expected from Eq.
(\ref{resolution}). The larger spot diameter measured at low
temperature is presumably due to change of the objective lens shape
caused by thermal contraction.

\section{Measurements and Discussion}

The operation of the Kerr microscope is demonstrated by measuring
the time-resolved Kerr rotation (TRKR) response of a GaAs/AlGaAs
heterojunction structure containing a two-dimensional electron gas
(2DEG). Detailed description of the material was given elsewhere
\cite{Pugzlys07}. Briefly, the substrate is a [001] oriented
\textit{i}-GaAs crystal. A multilayer buffer consisting of ten
periods of GaAs/AlAs grown on the substrate is used to smoothen the
substrate surface. A 9330 {\AA} layer of undoped GaAs, called the
accumulation layer, is grown over the multilayer buffer. A 368 {\AA}
spacer layer of undoped Al$_{32}$Ga$_{68}$As is deposited on top of
the accumulation layer. Next, a donor layer consisting of 719 {\AA}
of Si-doped Al$_{32}$Ga$_{68}$As with $\sim 1\times10^{18}$
dopants/cm$^{3}$ donates electrons that form the 2DEG at the
interface between the accumulation layer and the spacer layer. The
heterostructure is capped with 55 {\AA} of \textit{n}-GaAs.

\begin{figure}
\begin{center}
\includegraphics[width=7cm]{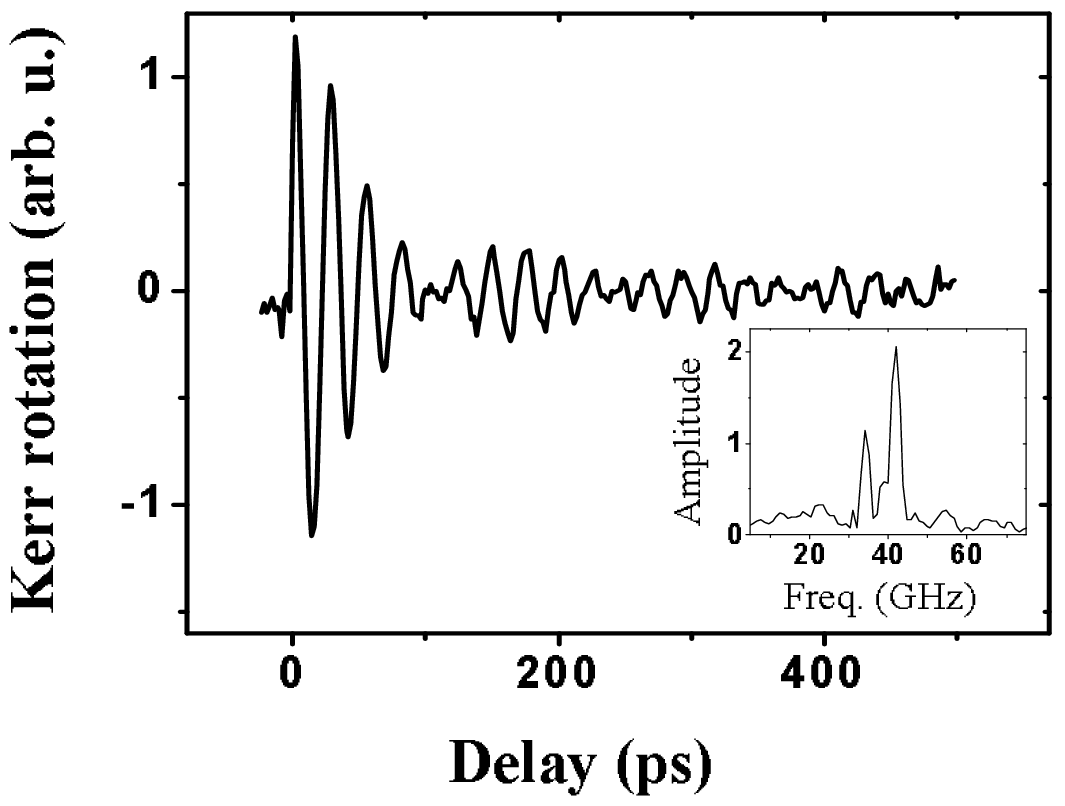}
\end{center}
\caption{Time-resolved Kerr rotation response of the heterojunction
2DEG sample measured with the Kerr microscope described in this
article. The vertical scale is the Kerr rotation angle and the
horizontal is the delay time between pump and probe pulses. The
oscillations of the Kerr angle are due to precession of spins in the
external magnetic field. The beating pattern of the Kerr signal
shows that at least two different population of spins contribute to
the total response. The inset is the Fourier transform of this data.
Two strong peaks, corresponding to the two populations observed, are
clearly seen in the Fourier spectrum. The low frequency peak
($|g^{\ast}_{1}|=0.34$) corresponds to electrons spins in the 2DEG
and the high frequency peak ($|g^{\ast}_{2}|=0.44$), to electron
spins in the underlying bulk GaAs layers. Data taken at 4.2 K and 7
Tesla with pump and probe spectra centered at 780 nm and 820 nm
respectively.} \label{TRKR}
\end{figure}

A typical TRKR trace measured with the microscope is shown in Fig.
\ref{TRKR}. The Kerr rotation angle is plotted in the vertical axis
as a function of pump and probe delay. This trace was taken at 4.2 K
and 7 Tesla with excitation density $1.1\times10^{12}$
photons/cm$^{2}$ per pump pulse. Pump and probe spots were
overlapped and had slightly larger diameters than the optimum values
measured before. Oscillations of the Kerr angle in the TRKR trace
are due to precession of the optically oriented spins in the applied
magnetic field. From the measured precession frequency $\nu$, the
effective g-factor $g^\ast$ of conduction electrons can be obtained
utilizing the relation:

\begin{equation}
\\g^\ast=h\nu/\mu_{B}\vec{B} \label{Larmorfreq}
\end{equation}

\noindent where $\mu_{B}$ is Bohr's magneton, $\vec{B}$ is the
applied magnetic field, and $h$ is Planck's constant. Fourier
transform of this TRKR measurement (inset in Fig. \ref{TRKR}) shows
that the signal is composed of two precession frequencies
corresponding to two different spin populations in the
heterostructure. Using the above equation, g-factor values of
$|g^{\ast}_{1}|=0.34$ and $|g^{\ast}_{2}|=0.44$ are found. These
g-factors correspond to electrons in the 2DEG and electrons in the
underlying bulk \textit{i}-GaAs layers \cite{Rizo08}.

The measured Kerr rotation signal decays as a function of time due
to several mechanisms. One of these mechanisms is the randomization
of the orientation of the transverse component of spins. In analogy
with nuclear magnetic resonance, this ensemble dephasing is
characterized by the time-constant $T^* _{2}$. Another mechanism
that influences the observed signal decay is electron-hole
recombination which is characterized by the time constant
$\tau_{e}$. Under certain conditions, a net spin polarization
remains even after electron-hole recombiation \cite{Kikkawa98}, in
which case $\tau_{e}$ can effectively be regarded as being infinite.
If this is not the case, the total decay time-constant of the Kerr
signal $\tau_{K}$ will be given by: $1/\tau_{K}=1/T^*
_{2}+1/\tau_{e}$. Additionally, the observed Kerr rotation signal
may decay due to diffusion of spin polarized electrons out of the
area covered by the focused probe beam. This mechanism is important
when the area of the focused probe beam is comparable to or less
than $D\cdot\tau_{K}$, where $D$ is the carrier diffusion constant.
This mechanism of signal decay must be taken into account when
measuring with tightly focused beams as is the case with the
microscope described here.

To separate the different contributions to the decay of the total
spin signal it is necessary to perform additional experiments.
Isolating the spin ($T^* _{2}$) and electronic ($\tau_{e}$)
contributions to $\tau_{K}$ requires independently measuring the
carrier dynamics. A transient reflectance experiment measures the
dynamics of photoexcited carriers irrespective of their spin
polarization from which $\tau_{e}$ can be determined. Changing from
a TRKR configuration to a transient reflectance configuration can be
done simply by replacing the photoelastic-modulator with an optical
chopper and measuring changes in the total reflected probe intensity
induced by excitation with a linearly polarized pump pulse. The
carrier diffusion constant $D$ can be determined by a transport
measurement or utilizing the ability of the microscope to excite
(spin polarized) carriers in one location and probe them a certain
distance away as will be discussed below.

A remarkable feature of the instrument described here is its ability
to measure transport of spin polarized electrons under high magnetic
fields. High magnetic fields are crucial, for example, in order to
distinguish between different electron populations with slightly
different values of $|g^{*}|$. This allows identifying the electron
populations that give rise to the signal in spin transport
measurements. Such is the case of the heterojunction sample
described above. The g-factors of electrons located in the 2DEG and
in the bulk layers differ only by a small amount. In order to
clearly distinguish between the two electron populations, the
difference in their precession frequencies must be comparable to or
larger than the Kerr signal decay rate of both spin species. At high
enough magnetic fields, the small difference in $|g^{*}|$ transforms
into a substantial difference in precession frequencies. In this
case it becomes possible to identify the contribution to the
observed transport of spin polarization from each electron
population.

\begin{figure}
\begin{center}
\includegraphics[width=7cm]{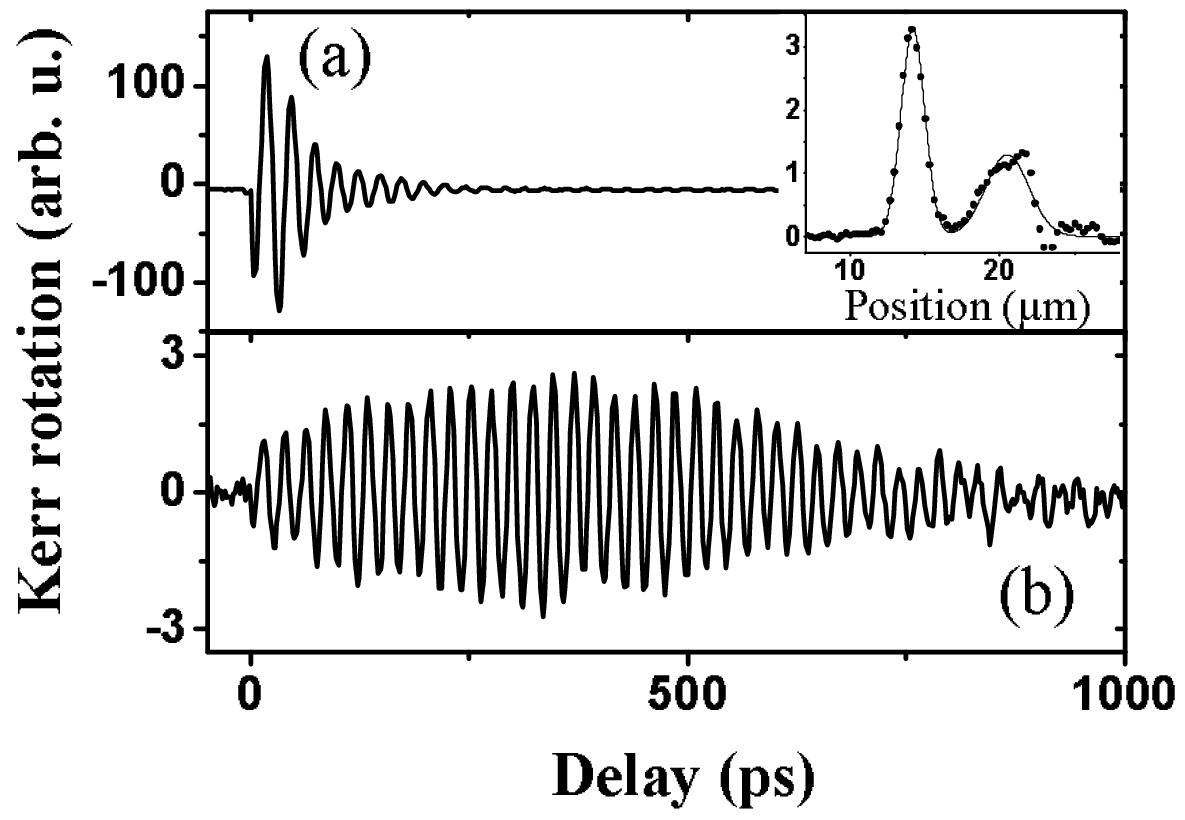}
\end{center}
\caption{Time-resolved Kerr rotation response of the heterojunction
2DEG sample at high photoexcitation density ($4.2\times10^{14}$
photons/cm$^{2}$). Plot (a) measured with overlapped pump and probe
spots shows a monotonic decay of the signal. Plot (b) measured with
pump and probe spots separated by 6 $\mu$m. The inset shows the
measured spatial profile (dots) and Gaussian fit (solid line) of
non-overlapped pump (right peak, FWHM $\sim$3.4 $\mu$m) and probe
(left peak, FWHM $\sim$1.9 $\mu$m) beams. In case (b), the Kerr
signal grows for the first few hundred picoseconds showing that
diffusion of spin polarized carriers takes place from the area
covered by the pump spot to the probed area. At delays greater than
approx. 400 ps the signal starts to decay due to spin dephasing,
electron-hole recombination and electron diffusion outside of the
probed area. Fourier transform of this Kerr rotation trace reveals a
single population with $|g^{\ast}_{2}|=0.44$. Thus, at this high
photoexcitation density, the spin diffusion signal is dominated by
carriers from the underlying bulk GaAs layers. Data taken at 4.2 K
and 7 Tesla with pump and probe spectra centered at 780 nm and 820
nm respectively.} \label{sixmicron}
\end{figure}

Spin polarization transport was studied in the heterojunction sample
in order to determine the nature of the carriers involved in spin
transport under high magnetic field and high photoexcitation
density. With overlapped pump and probe spots, the Kerr rotation
signal amplitude was first maximized by adjusting the focus. Figure
\ref{sixmicron}.a shows the TRKR signal obtained in this
configuration at 7 Tesla and 4.2 K. Subsequently, the pump spot was
displaced away from the probe spot. In this configuration, the spot
profiles and separation were measured by the knife edge technique
described earlier (inset in Fig. \ref{sixmicron}). Spot diameters
(FWHM) were measured to be $\sim$1.9 $\mu$m and $\sim$3.4 $\mu$m for
the pump and probe respectively. The difference in spot diameters is
due to the different central wavelengths of the two beams (780 nm
and 820 nm for the pump and probe respectively) and chromatic
aberration from the single-element objective lens. The separation
between the pump and probe spots is 6 $\mu$m. From the measured spot
diameters the pump photon density is determined to be
$4.2\times10^{14}$ photons/cm$^{2}$ while the probe photon density
is approximately half of this. This photon density is 3 orders of
magnitude higher than the density of electrons in the 2DEG.

The TRKR trace measured with 6 $\mu$m separation between the pump
and probe spots is shown in plot (b) of Fig. \ref{sixmicron}. With
overlapped pump and probe spots (plot (a) in Fig. \ref{sixmicron}),
the amplitude of the Kerr signal decays monotonically with time.
This is determined both by the decay of spin polarization and by the
diffusion of carriers away from the area covered by the pump and
probe beams. The spin polarized carriers that diffuse away from the
area covered by the pump are detected when measured with
non-overlapped spots. The time it takes for the carriers to diffuse
6 $\mu$m gives rise to the slow increase in the amplitude of the
envelope of the Kerr signal that peaks at a delay of approximately
400 ps. In addition, as will be discussed below, the Kerr signal
with separated pump and probe shows coherence up to much longer
delays. It is important to note that diffusion takes place
predominantly in the plane of the sample and not along the growth
direction because the multilayer buffer between the sample substrate
and accumulation layer acts as a barrier for carrier diffusion.
Fourier analysis of the data taken with non-overlapped spots reveal
that only electrons with $|g^{\ast}|=0.44$, that is bulk GaAs
electrons, are observed in this measurement.

The signal obtained with non-overlapped spots can be modeled
assuming that the width of the Gaussian profile of the carrier
density excited by the pump pulse increases with time due to carrier
diffusion with diffusion constant $D$. As spin polarized carriers
diffuse radially, some enter the region covered by the probe spot
increasing the amplitude of the Kerr signal. Spin relaxation also
takes place, which ultimately causes a decrease in the Kerr signal
amplitude. Analysis of the data reveals that the Kerr signal decay
time-constant $\tau_{K}$ is $\sim$700 ps and the diffusion constant
$D$ is $\sim$70 cm$^{2}$/s. This value of the diffusion constant is
consistent with the conclusion that only bulk GaAs spins contribute
to the observed signal since the diffusion constant of 2DEG
electrons is more than an order of magnitude larger.

Time-resolved Kerr rotation measured with large diameter ($>$100
$\mu$m) overlapped spots is insensitive to signal loss due to
diffusion of carriers out of the area covered by the laser spots.
Experiments with large spots on the same sample and under similar
conditions to those discussed above \cite{Rizo08} yield a $\tau_{K}$
no larger than 300 ps. This is more than two times shorter than the
$\tau_{K}$ retrieved from the measurements with non-overlapped
spots. The reason for the slower spin relaxation rate observed with
non-overlapped spots is probably that spin polarized electrons
become spatially separated from holes due to their larger diffusion
constant. Electron spin-flips due to electron-hole scattering has
been shown to be an important spin relaxation mechanism in GaAs
quantum-wells \cite{Wagner93, Gotoh00}.

The Kerr signal decay time constant of 2DEG electrons decreases as
the excitation density $N_{ex}$ increases\cite{Rizo08} approximately
as $\tau_{K}\propto N^{-0.4} _{ex}$. This is the reason that at high
photon densities only electrons with $g^{\ast}=-0.44$, i.e.,
electrons in bulk GaAs, contribute to the Kerr signal. At low
photoexcitation densities spin polarized electrons in the 2DEG can
be resolved in TRKR measurements as shown in Fig. \ref{TRKR}.

\begin{figure}[!h]
\begin{center}
\includegraphics[width=7cm]{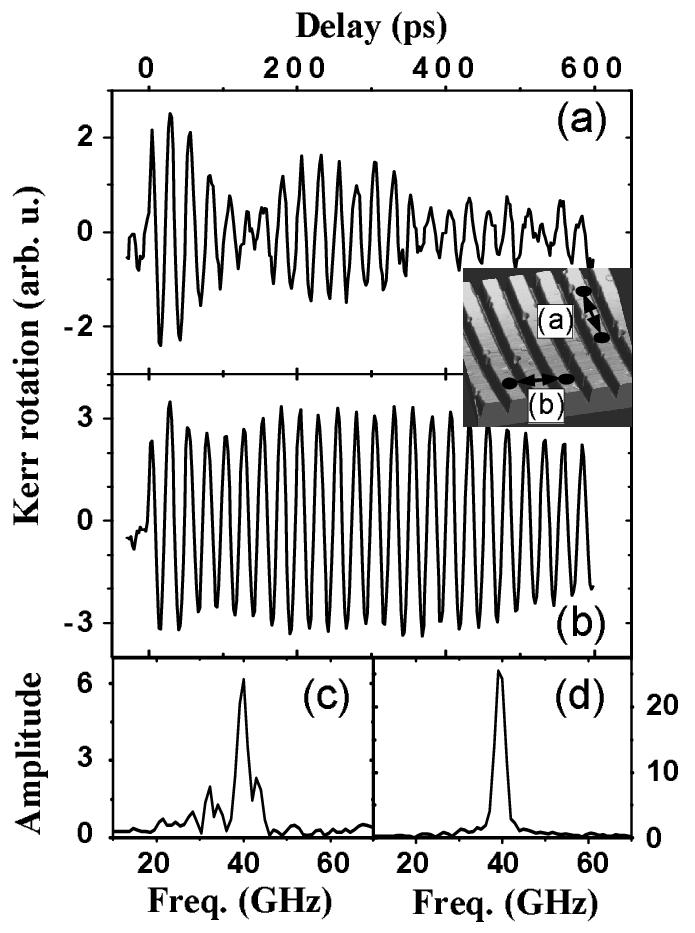}
\end{center}
\caption{Time-resolved Kerr rotation measurements on the 2DEG sample
with quasi-1D channels taken with pump and probe spots separated by
$\sim$4 $\mu$m. Plot (a) shows measurements with pump and probe
spots focused along the same channel as depicted in the inset. Two
spin populations contribute to the total Kerr signal as evidenced by
the beating pattern and by the Fourier spectrum of this data, shown
in plot (c). The high frequency population ($\sim$42 GHz)
corresponds to electrons in bulk GaAs. The low frequency population
($\sim$32 GHz) corresponds to electrons in the 2DEG that diffuse
along the quasi-1D channel from the area covered by the pump spot to
the probed area. Plot (b) shows measurements with pump and probe
spots separated by $\sim$4 $\mu$m but focused on different channels
in the manner depicted in the inset. Fourier transform of this
trace, plot (d), presents a single peak corresponding to electrons
with $g^{\ast}=-0.44$. Electron diffusion through the 2DEG is
inhibited by the ridges defining the quasi-1D structures. Thus, only
electrons diffusing from the area covered by the pump spot through
the underlying bulk GaAs reach the probed area. Data taken at low
photon-density ($1\times10^{12}$ photons/cm$^{2}$), 80 K and 7 Tesla
with pump and probe spectra centered at 780 nm and 820 nm
respectively.} \label{alongacross}
\end{figure}

Low excitation density ($1\times10^{12}$ photons/cm$^{2}$)
measurements were realized on a single quasi-1D wire fabricated on
the heterojunction sample and demonstrate spin transport by 2DEG
carriers. A grating structure with 1.6 $\mu$m period was fabricated
on the sample surface by lithography and wet chemical etching. 200
nm wide and 120 nm deep ridges defined channels on the 2DEG of 1.4
$\mu$m width and 180 $\mu$m length. Spin transport along one of
these channels was measured by TRKR with pump and probe spots
separated by $\sim$4 $\mu$m. The result is shown in Fig.
\ref{alongacross}.a. The Fourier spectrum of this trace shows that
two spin populations contribute to the Kerr signal in this
configuration: electrons in bulk GaAs and electrons in the 2DEG
(peaks at 42 GHz and 32 GHz respectively in Fig.
\ref{alongacross}.c) \cite{Rizo08}. Bulk electrons diffuse
isotropically in the plane normal to the growth direction. Quasi-1D
electrons, on the other hand, diffuse only along the channel because
the ridges of the grating confine the carriers to remain within the
channel. For this reason, when the probe spot is focused on a
different channel than the pump pulse, no Kerr response from the
2DEG is observed. Figure \ref{alongacross}.b shows the TRKR signal
measured with pump and probe spots separated by $\sim$4 $\mu$m but
each one focused on a different channel. The Fourier spectrum of
this measurement, Fig. \ref{alongacross}.d, shows no peak at the
precession frequency of the 2DEG spins. This experiment demonstrates
that the Kerr microscope can measure transport of spin polarization
at distances as short as 4 $\mu$m. The ability to resolve spin
transport with such high spatial resolution is essential, for
example, for measuring ballistic spin transport where the relevant
length scale is of the order of the electron mean free path.

\section{Summary}

In this article a compact cryogenic sample-scanning microscope
designed for static and time-resolved Kerr rotation measurements in
high magnetic fields is described. The instrument has spatial
resolution of $\sim$1 $\mu$m and temporal resolution utilizing the
full laser spectrum of 20 fs. The microscope features the capability
of easily changing from Voigt configuration to Faraday configuration
due to the use of compact optics and translation systems on the
insert of a magneto-optical cryostat with split-coil magnet. The
instrument is designed for measuring spin transport in micro and
nanostructures and can thus polarize and probe spins at different
locations on a device.

The capabilities of the microscope were demonstrated by measuring
transport of spin polarization by quasi-1D electrons over micrometer
distances such that we can now explore ballistic transport. For
these experiments high spatial resolution, high magnetic field
operation and the ability to spatially separate pump and probe spots
on the sample are necessary.

The power and versatility of the Kerr microscope described above
anticipate diverse applications in spintronics such as tracing of
spin packet transport in spintronic circuits, investigation of
spin-filtering of electric currents in interconnected wire
structures \cite{Kiselev01,Yamamoto05}, mapping the effect of
anisotropic spin-orbit fields on spin diffusion in 2D electron
systems \cite{Meier07} and characterization of ballistic spin
transport.

\section{Acknowledgements}

This work is supported by the Zernike Institute for Advanced
Materials, by the Dutch Foundation for Fundamental Research on
Matter (FOM) and by The Netherlands Organization for Scientific
Research (NWO). Two of us (D.R. and A.D.W) acknowledge gratefully
financial support of the Deutsche Forschungsgemeinschaft within the
SFB491. Special thanks to H. van Driel, U. Douma and A. Kamp from
the University of Groningen for the design and fabrication of
critical components of the microscope and the controlling and
measuring software.


\begin{thebibliography}{99}

\bibitem{Kikkawa99} J. M. Kikkawa and D. D. Awschalom
\textit{Nature} \textbf{397}, 139 (1999).

\bibitem{Beach05} G. S. D. Beach, C. Nistor, C. Knutson, M. Tsoi and J. L. Erskine
\textit{Nat. Mater.} \textbf{4}, 741 (2005).

\bibitem{Nistor06} C. Nistor, G. S. D. Beach and J. L. Erskine
\textit{Rev. Sci. Instrum.} \textbf{77}, 103901 (2006).

\bibitem{Atature07} M. Atature, J. Dreiser, A. Badolato, and A. Imamoglu
\textit{Nat. Phys.} \textbf{3}, 101 (2007).

\bibitem{Lo08} F.-Y. Lo, A. Melnikov, D. Reuter, Y. Cordier, and A. D. Wieck
\textit{Appl. Phys. Lett.} \textbf{92}, 112111 (2008).

\bibitem{Meyer05} C. Meyer, O. Sqalli, H. Lorenz and K. Karrai
\textit{Rev. Sci. Instrum.} \textbf{76}, 063706 (2005).

\bibitem{Pugzlys07} A. Pug\v{z}lys, P. J. Rizo, K. Ivanin, A. Slachter, D. Reuter, A. D. Wieck, C. H. van der Wal and
P. H. M. van Loosdrecht \textit{J. Phys.: Condens. Matter.}
\textbf{19}, 295206 (2007).

\bibitem{Rizo08} P. J. Rizo, A. Pug\v{z}lys, A. Slachter, D. Reuter, A. D. Wieck, P. H. M. van Loosdrecht and C. H. van der Wal
\textit{In preparation}.

\bibitem{Kikkawa98} J. M. Kikkawa and D. D. Awschalom
\textit{Phys. Rev. Lett.} \textbf{80}, 4313 (1998).

\bibitem{Wagner93} J. Wagner, H. Schneider, D. Richards, A. Fischer and K. Ploog
\textit{Phys. Rev. B} \textbf{47}, 4786 (1993).

\bibitem{Gotoh00} H. Gotoh, H. Ando, T. Sogawa, H. Kamada T. Kagawa and H. Iwamura
\textit{J. Appl. Phys.} \textbf{87}, 3394 (2000).

\bibitem{Kiselev01} A. A. Kiselev and K. W. Kim
\textit{Appl. Phys. Lett.} \textbf{78}, 775 (2001).

\bibitem{Yamamoto05} M. Yamamoto, T. Ohtsuki and B. Kramer
\textit{Phys. Rev. B} \textbf{72}, 115321  (2005).

\bibitem{Meier07} L. Meier, G. Salis, I. Shorubalko, E. Gini, S. Sch\"{o}n and K. Ensslin
\textit{Nat. Phys.} \textbf{3}, 650 (2007).


\end{thebibliography}
\end{document}